\documentclass[reprint]{revtex4}

\usepackage{amssymb}
\usepackage{amsmath}

\usepackage{graphicx}
\usepackage[export]{adjustbox}

\usepackage{xparse}

\usepackage{hyperref}
\usepackage{hypernat}

\usepackage[caption=false,labelfont=normalsize,labelformat=empty,textfont=normalsize,justification=centering]{subfig}

\renewcommand{\Re}{\mathrm{Re}}

\newcommand{\iu}{\mathrm{i}}

\begin{document}

\title{Mass-derivative relations for leptogenesis}

\author{Tom\'a\v s Bla\v zek}
\email{tomas.blazek@fmph.uniba.sk}
\author{Peter Mat\'ak}
\email{peter.matak@fmph.uniba.sk}
\affiliation{Department of Theoretical Physics, Comenius University,\\ Mlynsk\'a dolina, 84248 Bratislava, Slovak Republic}

\date{\today}

\begin{abstract}
In this work, a diagrammatic representation of thermal mass effects is derived from the $S$-matrix unitarity both in the classical and quantum Boltzmann equations. Within the example of the seesaw type-I leptogenesis, we discuss the connection of the Higgs thermal mass and the cancelations of infrared divergences in zero- and finite-tem\-pe\-ra\-tu\-re calculations of the right-handed neutrino decay and scattering processes.
\end{abstract}

\maketitle

\section{Introduction}\label{sec1}

Quantum thermal corrections to reaction rates of the seesaw type-I leptogenesis at higher perturbative orders have been investigated in both the real-time \cite{Giudice:2003jh,Salvio:2011sf,Biondini:2013xua, Biondini:2015gyw, Biondini:2016arl} and ima\-gi\-na\-ry-time \cite{Laine:2011pq,Bodeker:2017deo,Bodeker:2019rvr} formalisms. A more complete treatment of the zero-temperature amplitudes can be found in Ref. \cite{Racker:2018tzw}, where interference between connected and disconnected amplitudes was taken into account. Here we fill in the gap between the two approaches. \emph{Completing the zero-tem\-pe\-ra\-tu\-re reaction rates at a given perturbative order we show that some of the contributions can be interpreted in terms of the finite-temperature field theory}.

The $S$-matrix unitarity and Cutkosky cutting rules are often used to describe $CP$ violation effects, or, via Ki\-no\-shi\-ta-Lee-Nauen\-berg theorem \cite{Kinoshita:1962ur, Lee:1964is, Frye:2018xjj}, the infrared finiteness in particles' interactions. Models of matter asymmetry generation in the early universe often contain massless particles leading to infrared divergences in reaction rates at higher perturbative orders. Here we focus on a simple extension of the Standard Model containing heavy right-handed neutrinos $N_i$ coupled to the Standard Model left-handed leptons and Higgs doublet via Yukawa interaction
\begin{align}\label{eq1}
\mathcal{L}\supset -\frac{1}{2}M_i\bar{N}_iN_i-\left(\mathcal{Y}_{\alpha i}\bar{N}_i P_L l_{\alpha}H + \mathrm{H.c.}\right)
\end{align}
with $M_i$ denoting the right-handed neutrino Majorana masses. The Higgs particles and standard model leptons are considered with zero nonthermal mass. When higher-order corrections due to the top Yukawa coupling \cite{Nardi:2007jp,Pilaftsis:2003gt,Pilaftsis:2005rv}, or gauge interactions \cite{Fong:2010bh} are included, some authors put the Higgs thermal mass in by hand to obtain infrared-finite results\footnote{Accuracy of this procedure as a higher-order correction to the right-handed neutrino decay rate has been discussed in Refs. \cite{Kiessig:2009cm,Kiessig:2010zz}.}. More recently, it has been argued by the Ki\-no\-shi\-ta-Lee-Na\-uen\-berg theorem that $2\leftrightarrow 3$ processes, in which connected and disconnected amplitudes interfere, may be used to cancel infrared divergences in the $2\leftrightarrow 2$ scatterings \cite{Racker:2018tzw}. In our work, the diagrammatic representation developed in Refs. \cite{Blazek:2021olf, Blazek:2021zoj} is used to study the connection of the two app\-roa\-ches to thermal masses and infrared finiteness at a given perturbative order. It is shown that some of the $2\leftrightarrow 3$ contributions are directly related to the Higgs thermal mass effects in the $N_i\rightarrow lH$ decay kinematics. The mass-derivative formula has originally been derived for the propagator in thermal field theory in Ref. \cite{Fujimoto:1984kh}. Although this paper is not recent, here, for the first time, we derive similar type of relations for the reaction rates in the classical Boltzmann equation. As the next step, we confirm their validity for quantum thermal-corrected rates. The results are important as they are applicable to other models containing forward diagrams similar to those considered in this paper.

The article is structured as follows. In section \ref{sec2} we consider the simplest in\-fra\-red-finite example of the mass-de\-ri\-va\-ti\-ve relation, originating from the Higgs self-coupling correction to the $N_i\rightarrow lH$ decay rate. There the classical thermal mass is obtained from the Maxwell-Boltzmann statistics. Section \ref{sec3} deals with the inclusion of quantum thermal corrections using the cylindrical diagrammatic representation of Ref. \cite{Blazek:2021zoj}. Finally, in section \ref{sec4}, the top Yukawa corrections, where nontrivial wave-function renormalization of the Higgs field occurs, are studied.

\section{Thermal mass effects in zero-temperature calculations}\label{sec2}

At the leading perturbative order, the momentum-integrated Boltzmann equation for the right-handed neutrino number density, in the universe expanding at the Hubble rate $\mathcal{H}$, can be written as
\begin{align}\label{eq2}
\dot{n}_{N_i\vphantom{\bar{H}}}+3\mathcal{H}n_{N_i\vphantom{\bar{H}}}= -\mathring \gamma_{N_i\rightarrow lH\vphantom{\bar{l}\bar{H}}}-\mathring \gamma_{N_i\rightarrow \bar{l}\bar{H}}+\mathring \gamma_{lH\vphantom{\bar{l}\bar{H}}\rightarrow N_i}+\mathring \gamma_{\bar{l}\bar{H}\rightarrow N_i}
\end{align}
The small circle indicates the decay and inverse decay reaction rates calculated using \emph{zero-temperature} quantum field theory and \emph{classical} Maxwell-Boltzmann phase-space densities denoted $\mathring{f}$. We shall discuss the inclusion of quantum statistics in the next section. At temperatures of the order of the right-handed neutrino mass, Higgs particles are effectively massless. However, for a while, let us consider the reaction rate with finite Higgs mass. Then
\begin{align}\label{eq3}
\mathring\gamma_{N_i\rightarrow lH\vphantom{\bar{l}\bar{H}}}=
\frac{1}{4\pi}(\mathcal{Y}^\dagger\mathcal{Y})_{ii}M^2_i
\left(1-\frac{m^2_H}{M^2_i}\right)^2\int[d\mathbf{p}_{N_i}]\mathring{f}_{N_i}
\end{align}
where $[d\mathbf{p}_{N_i}]=d^3\mathbf{p}_{N_i}/((2\pi)^3 2E_{N_i})$. Throughout this work, kinetic equilibrium is assumed for all particle species, with the right-handed neutrinos out of chemical equilibrium\footnote{Concerning the assumption of the right-handed neutrinos being in kinetic equilibrium, see Ref. \cite{Hahn-Woernle:2009jyb} for a detailed discussion.}. We will further denote the particles' four-momenta by a subscript (such as $p_{N_i}$), while the four-momentum of the internal Higgs lines will be labeled simply by $k$. The energy dependence of the phase-space densities will only be explicit for the internal Higgs particles or whenever necessary. 

\begin{figure}[t!]
\subfloat{\label{fig1a}}
\subfloat{\label{fig1b}}
\centering\includegraphics[scale=1]{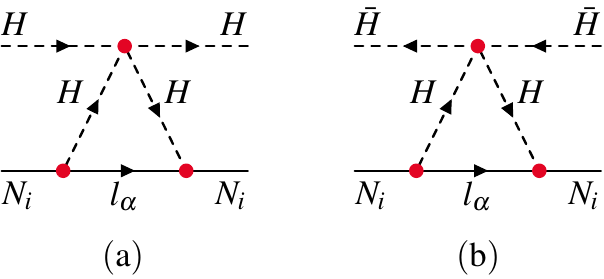}
\caption{\label{fig1} Forward scattering diagrams with the $N_iH$ and $N_i\bar{H}$ initial states at the $\mathcal{O}(\mathcal{Y}^2\lambda)$ order.}
\end{figure}

Now, one may ask if any correction to the right-hand side of Eq. \ref{eq2} occurs when the first-order in the Higgs self-coupling from $\mathcal{L}\supset -\lambda\left(H^\dagger H\right)^2$
is taken into account. The positive answer is rooted in the fact that any contribution to the reaction rate can be obtained from the thermal average of a square of amplitude and thus by cutting a forward diagram. Instead of the standard Cutkosky cuts, we use the $S$-matrix unitarity condition directly for $S=1+\iu T$, simplifying the treatment of on-shell intermediate states. From
\begin{align}\label{eq4}
1-\iu T^\dagger = (1+\iu T)^{-1}\quad\rightarrow\quad \iu T^\dagger = \iu T - (\iu T)^2 +  \ldots
\end{align}
we receive for the amplitude squared the relation \cite{Blazek:2021olf}
\begin{align}\label{eq5}
\vert T_{fi}\vert^2 = -\iu T_{if}\iu T_{fi} + \sum_n \iu T_{in}\iu T_{nf}\iu T_{fi} -  \ldots
\end{align}
Therefore, each contribution to a reaction rate for a particular process can be obtained from a forward diagram with one or more on-shell intermediate states included with appropriate sign according to Eq. \eqref{eq5}.  The simplest nontrivial forward diagrams at the $\mathcal{O}(\mathcal{Y}^2\lambda)$ order, containing $N_i$ in the initial state, can be seen in Fig. \ref{fig1}. Cutting the diagram in Fig. \ref{fig1a} then leads to
\begin{align}\label{eq6}
\mathring{\gamma}_{N_iH\rightarrow lHH} = \hskip1.5mm
-\hskip1.5mm\includegraphics[scale=1, valign=c]{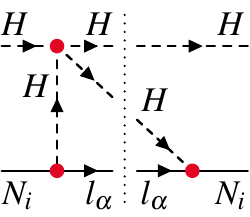}
\hskip1.5mm-\hskip1.5mm\includegraphics[scale=1, valign=c]{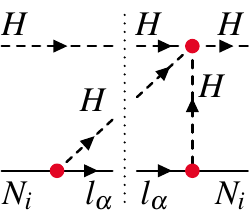}
\hskip1.5mm+\hskip1.5mm\includegraphics[scale=1, valign=c]{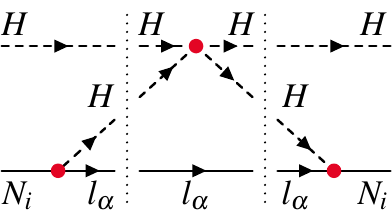}
\end{align}
that represents the rate of the $N_iH\rightarrow lHH$ reaction. Each of the on-shell cuts indicated by the dotted lines corresponds to the phase-space integration with a four-momentum conservation delta function. Thus, for the cut in the middle of the first term in Eq. \eqref{eq6} we have
\begin{align}\label{eq7}
\int[d\mathbf{p}_{l}][d\mathbf{k}][d\mathbf{k}'](2\pi)^4\delta^{(4)}(p_{N_i}+p_H -p_l-k-k')
\end{align}
while the standalone Higgs line in the disconnected part brings $(2\pi)^3 2E_H\delta^{(3)}(\mathbf{p}_{H}-\mathbf{k}')$,
such that the $[d\mathbf{k}']$ integration can be done immediately putting $p^{\vphantom{'}}_H=k'$. Furthermore, the phase-space integration with circled densities of the initial state particles,
\begin{align}\label{eq8}
\int[d\mathbf{p}_{N_i}][d\mathbf{p}_{H}]\mathring{f}_{N_i}\mathring{f}_H
\end{align}
has to be included in Eq. \eqref{eq6}. The summation over the discrete degrees of freedom is always implicit. In the first and second terms in Eq. \eqref{eq6}, the uncut Higgs propagators come with the same four-mo\-men\-tum $k$ as the on-shell Higgs in the intermediate state, leading to a singular expression. A similar situation has been encountered in different processes in Refs. \cite{Frye:2018xjj, Racker:2018tzw}, using standard Cutkosky rules. Here, to cure the singularity, we employ
\begin{align}\label{eq9}
\frac{\iu}{k^2+\iu\epsilon} = \mathrm{P.V.}\frac{\iu}{k^2}+\pi\delta(k^2)
\end{align}
diagrammatically expressed as \cite{Blazek:2021olf}
\begin{align}\label{eq10}
\includegraphics[scale=1, valign=c]{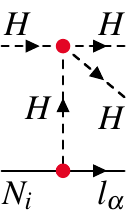}\hskip1.5mm = 
\hskip1.5mm\mathrm{P.V.}\includegraphics[scale=1, valign=c]{math2a.pdf}\hskip1.5mm + 
\hskip1.5mm\frac{1}{2}\includegraphics[scale=1, valign=c]{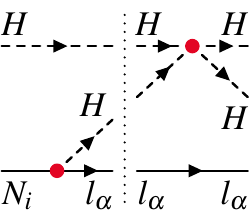}
\end{align}
to replace the Higgs internal lines. Then the contribution of the last double-cut diagram in Eq. \eqref{eq6}, containing the square of the delta function, is completely canceled by what is obtained from the first two diagrams when using Eq. \eqref{eq10}. To evaluate the reaction rate $\mathring{\gamma}_{N_iH\rightarrow lHH}$, we need to integrate over the intermediate-state phase-space. For that purpose, we use the identity \cite{Fujimoto:1984kh}
\begin{align}\label{eq11}
2\delta_+(k^2)\mathrm{P.V.}\frac{1}{k^2} = -\frac{1}{(k^0+\vert\mathbf{k}\vert)^2}\frac{\partial\delta(k^0-\vert\mathbf{k}\vert)}{\partial k^0}.
\end{align}
where $\delta_+(k^2)=\theta(k^0)\delta(k^2)$. Summation over the spin, iso\-spin, and flavor then gives
\begin{align}\label{eq12}
\mathring{\gamma}_{N_iH\rightarrow lHH}=
-\frac{3}{\pi}\lambda(\mathcal{Y}^\dagger\mathcal{Y})_{ii}\int[d\mathbf{p}_{N_i}][d\mathbf{p}_H]\mathring{f}_{N_i}\mathring{f}_H.
\end{align}
Stran\-ge\-ly enough, the expression in Eq. \eqref{eq12} is negative. This would mean that even though the right-handed neutrino $N_i$ appears in the initial state and not in the final state, its abundance is increased in this process. Instead, we rather suggest taking this reaction rate as a mere correction to the $N_i\rightarrow lH$ decay, the process already included in Eq. \eqref{eq2} at the leading order. To see that this is indeed the case, let us consider the Higgs thermal mass due to its self-interaction in a thermal medium, defined as
\begin{align}\label{eq13}
\mathring{m}^2_{H,\lambda}(T)=12\lambda\int[d\mathbf{p}_H]\mathring{f}_H=\frac{3}{\pi^2}\lambda T^2,
\end{align} 
where, however, the usual Bose-Einstein density has been replaced by $\mathring{f}_H=\exp \left\{-E_H/T\right\}$ to treat the reaction rates consistently. Otherwise, a more common result of an uncircled thermal mass $m^2_{H,\lambda}(T)=\frac{1}{2}\lambda T^2$ would be obtained \cite{Giudice:2003jh}. Taking Eqs. \eqref{eq3} and \eqref{eq12} into consideration, we can easily check that
\begin{align}\label{eq14}
\mathring{\gamma}_{N_iH\rightarrow lHH\vphantom{\bar{H}}}+\mathring{\gamma}_{N_i\bar{H}\rightarrow lH\bar{H}}=
\mathring{m}^2_{H,\lambda}(T)\frac{\partial}{\partial m^2_H} \bigg\vert_{m^2_H=0}
\mathring{\gamma}_{N_i\rightarrow lH}.
\end{align}
The contribution of the $N_i\bar{H}\rightarrow lH\bar{H}$ reaction originating from the cuttings of Fig. \ref{fig1b} is the same as in Eq. \eqref{eq12}. In Eq. \eqref{eq14}, we can observe a novel type of relation between the reaction rates of $2\rightarrow 3$ and $1\rightarrow 2$ processes. The former, given by Eq. \eqref{eq6}, thus may be understood as a diagrammatic representation of the Higgs thermal mass effect in the right-handed neutrino decay rate that, to our best knowledge, was not previously formulated in the literature.

\section{Mass-derivative and thermal corrections}\label{sec3}

\begin{figure}[t!]
\subfloat{\label{fig2a}}
\subfloat{\label{fig2b}}
\subfloat{\label{fig2c}}
\subfloat{\label{fig2d}}
\centering\includegraphics[scale=1]{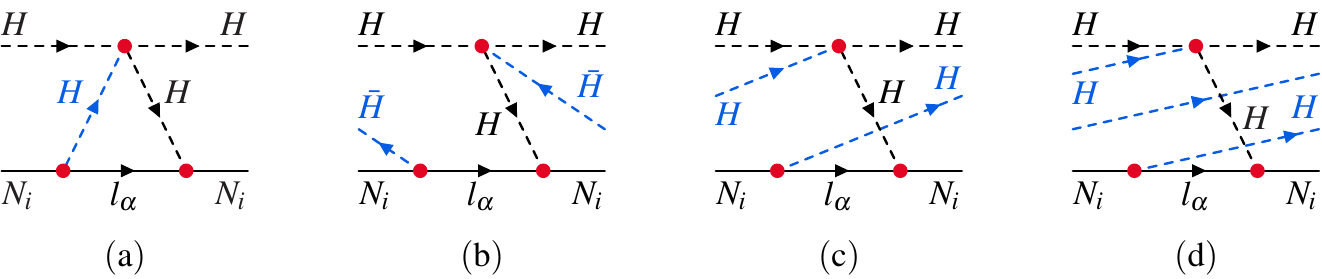}
\caption{\label{fig2} Forward scattering processes representing the contribution of the Higgs thermal propagator expanded into infinite series of zero-temperature diagrams.}
\end{figure}

Before proceeding to a more general case of the top Yu\-ka\-wa corrections in the next section, here we briefly discuss quantum thermal corrections to the mass-derivative relation in Eq. \eqref{eq14}. In Fig. \ref{fig1}, we considered the simplest forward diagrams at the $\mathcal{O}(\mathcal{Y}^2\lambda)$ order contributing to the Boltzmann equation for the $N_i$ density. However, at the same order in couplings, there are infinitely many forward diagrams, such as those in Fig. \ref{fig2}. The diagram in Fig. \ref{fig2b} can only be cut into a contribution to the rate of the forward $N_i\bar{H}H\rightarrow N_i\bar{H}H$ reaction. This does not change particle numbers and thus will be ignored. The diagrams in Figs. \ref{fig2c} and \ref{fig2d} can be cut in the same way as we observed in Eq. \eqref{eq6}. As discussed in our previous work \cite{Blazek:2021zoj}, drawing these diagrams on a cylindrical surface corresponds to the winding of the Higgs internal line in Fig. \ref{fig2a}. Summing over all winding numbers is equivalent to the replacement
\begin{align}\label{eq15}
\frac{\iu}{k^2+\iu\epsilon}\quad\rightarrow\quad\frac{\iu}{k^2+\iu\epsilon}+2\pi\sum^\infty_{w=1} [\mathring{f}_H(k^0)]^w\delta_+(k^2).
\end{align}
In thermal equilibrium, $\mathring{f}_H = \exp\left\{-k^0/T\right\}$ and the sum in Eq. \eqref{eq15} leads to Bose-Einstein distribution
\begin{align}\label{eq16}
f_H(k^0) = \sum^\infty_{w=1} [\mathring{f}_H(k^0)]^w = \frac{1}{\exp\left\{k^0/T\right\}-1}
\end{align}
reproducing the positive frequency part of the thermal propagator\footnote{In a general nonequilibrium case, it may seem unclear what the classical density $\mathring{f}_H$ is, while $f_H$ corresponds to the mean occupation number of a particular single-particle state in the Fock space. Nevertheless, for bosonic particles, one may formally define
$\mathring{f}_H = f_H/(1+f_H)$ and proceed along the same lines \cite{Blazek:2021zoj}.}. To account for thermal corrections to the reaction rate in Eq. \eqref{eq12} in a consistent way, the procedure introduced in Ref. \cite{Blazek:2021zoj} suggests the following. We should start with the diagram with the lowest winding numbers for all particles (the diagram in Fig. \ref{fig2a}) and consider all possible cuttings (such as in Eq. \eqref{eq6}). Then in each term, we include $f_{N_i}$ and $f_H$, the Fermi-Dirac and Bose-Einstein distributions, respectively, for the initial state particles. Similarly, the on-shell intermediate states (indicated by the dotted lines in Eq. \eqref{eq6}) receive factors $1-f_l$ and $1+f_H(k^0)$. This procedure is equivalent to the summation over the windings of the external and internal lines of the forward diagram before its cutting. Applying Eq. \eqref{eq9} to the uncut Higgs lines for each particular set of winding numbers, we obtain
\begin{align}\label{eq17}
\gamma_{N_iH\rightarrow lHH}=12\lambda\int [d\mathbf{p}_{N_i}][d\mathbf{p}_H] f_{N_i}f_H
\int d^4k \mathcal{F}(k^0,\mathbf{k})\;2\delta_+(k^2)\mathrm{P.V.}\frac{1}{k^2}
\end{align}
where
\begin{align}\label{eq18}
\mathcal{F}(k^0,\mathbf{k}) = \frac{1}{4\pi^2}(\mathcal{Y}^\dagger\mathcal{Y})_{ii}(M^2_i-k^2)\delta_+[(p_{N_i}-k)^2]
[1+f_H(k^0)][1-f_l(E_{N_i}-k^0)].
\end{align}
Then, using Eq. \eqref{eq11} and replacing $\mathring{f}_H$ by $f_H$ in Eq. \eqref{eq13}, we can write
\begin{align}\label{eq19}
\gamma_{N_iH\rightarrow lHH}=m^2_{H,\lambda}(T)\int [d\mathbf{p}_{N_i}]f_{N_i}
\int d^3\mathbf{k}\frac{\partial}{\partial k^0}\bigg\vert_{k^0=\vert\mathbf{k}\vert} \frac{\mathcal{F}(k^0,\mathbf{k})}{(k^0+\vert\mathbf{k}\vert)^2}.
\end{align}
The $N_i\rightarrow lH$ decay rate, on the other hand, with quantum statistical factors for the final states, equals
\begin{align}\label{eq20}
\gamma_{N_i\rightarrow lH}=
\int [d\mathbf{p}_{N_i}]f_{N_i}
\int \frac{d^3\mathbf{k}}{2E_{\mathbf{k}}}\mathcal{F}(E_{\mathbf{k}},\mathbf{k})
\end{align}
for $E_{\mathbf{k}}=\sqrt{\mathbf{k}^2+m^2_H}$ and one can easily check that
\begin{align}\label{eq21}
\frac{\partial}{\partial k^0}\bigg\vert_{k^0=\vert\mathbf{k}\vert} \frac{\mathcal{F}(k^0,\mathbf{k})}{(k^0+\vert\mathbf{k}\vert)^2}=
\frac{\partial}{\partial m^2_H} \bigg\vert_{m_H=0}\frac{\mathcal{F}(E_{\mathbf{k}},\mathbf{k})}{2E_{\mathbf{k}}}.
\end{align}
Comparing Eqs. \eqref{eq19} and \eqref{eq20}, we obtain directly the same formula as in Eq. \eqref{eq14} with a significant difference - the mass-derivative relation holds for the reaction rates containing complete quantum statistics and thermal mass.

\section{Mass-derivative and infrared finiteness}\label{sec4}

\begin{figure}[t!]
\subfloat{\label{fig3a}}
\subfloat{\label{fig3b}}
\subfloat{\label{fig3c}}
\subfloat{\label{fig3d}}
\centering\includegraphics[scale=1]{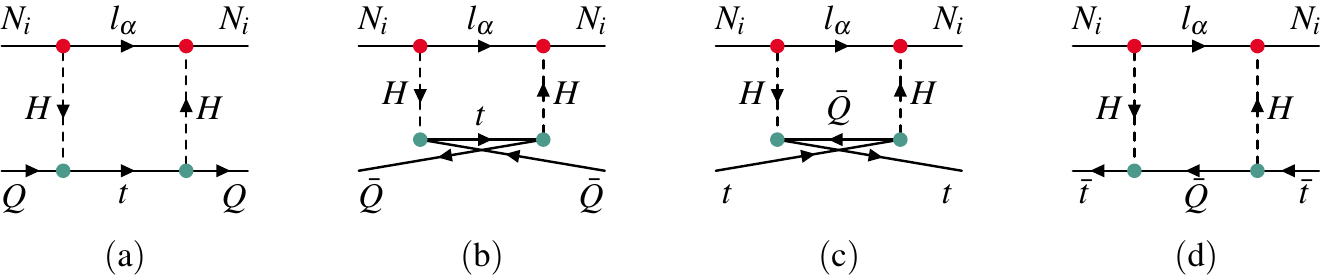}
\caption{\label{fig3} Forward diagrams for the $\mathcal{O}(\mathcal{Y}^2\mathcal{Y}^2_t)$ corrections to the $N_i\rightarrow lH$ decay rate and the corresponding neutrino quark scatterings. Similar diagrams, with all arrows reversed, contribute to $N_i\rightarrow \bar{l}\bar{H}$.}
\end{figure}

In addition to the Lagrangian density in Eq. \eqref{eq1}, let us consider the Yukawa interactions of the left-handed quark doublet $Q$, right-handed top quark $t$, and the Higgs field specified by
\begin{align}\label{eq22}
\mathcal{L}\supset -\mathcal{Y}_t \bar{t} P_L Q H + \mathrm{H.c.}.
\end{align}
The evolution of the right-handed neutrino density in Eq. \eqref{eq2} then receives corrections of the order $\mathcal{O}(\mathcal{Y}^2\mathcal{Y}^2_t)$ coming from various reactions. Here we focus on an example of the $N_i$ scattering with quark or antiquark. Then the corresponding forward diagrams shown in Fig. \ref{fig3} can be cut in two ways. For the diagram in Fig. \ref{fig3a}, one of them includes cutting the lepton and top quark internal lines,
\begin{align}\label{eq23}
\mathring{\gamma}_{N_iQ\rightarrow lt}= \hskip1.5mm-\hskip1.5mm \includegraphics[scale=1, valign=c]{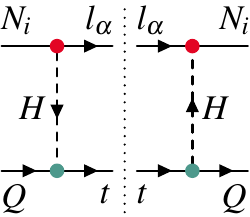}
\end{align}
that, in the case of massless quarks, contains an infrared divergence when $Q$ and $t$ momenta are collinear. To regularize it, we follow the literature \cite{Racker:2018tzw} and introduce a small mass $m$ to these particles. For the reaction rate, we obtain
\begin{align}\label{eq24}
\mathring{\gamma}_{N_iQ\rightarrow lt}=& 6\mathcal{Y}^2_t(\mathcal{Y}^\dagger\mathcal{Y})_{ii}
\int[d\mathbf{p}_{N_i}][d\mathbf{p}_{Q}]\mathring{f}_{N_i}\mathring{f}_Q\\
&\times\int[d\mathbf{p}_{l}][d\mathbf{p}_{t}](2\pi)^4\delta^{(4)}(p_{N_i}+p_Q-p_l-p_t) 2p_{N_i}.p_l\frac{2p_Q.p_t}{[(p_Q-p_t)^2]^2}.\nonumber
\end{align}
where the integration over the $l$ and $t$ phase-space can be performed easily using the Lorentz invariance, leading to
\begin{align}\label{eq25}
\frac{1}{4\pi}\bigg[\frac{1}{2}+\frac{s}{M^2_i}- \frac{M^2_i}{s-M^2_i}\left(\ln\frac{M^2_i}{s-M^2_i}+\ln\frac{m}{\sqrt{s}}\right)\bigg]
\end{align}
with $s=(p_{N_i}+p_Q)^2$. The logarithm $\ln m/\sqrt{s}$ represents the infrared divergence that should be canceled by the contribution of other reactions.

\subsection{Infrared finiteness and the Ki\-no\-shi\-ta-Lee-Nauen\-berg theorem}

The diagram in Fig. \ref{fig3a} may also be cut in analogy to what we have seen in our first example in section \ref{sec2}, obtaining
\begin{align}\label{eq26}
\mathring{\gamma}_{N_iQ\rightarrow lHQ}=\hskip1.5mm
-\hskip1.5mm\includegraphics[scale=1, valign=c]{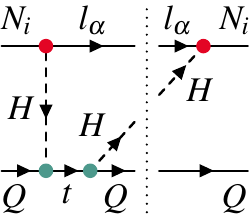}
\hskip1.5mm-\hskip1.5mm\includegraphics[scale=1, valign=c]{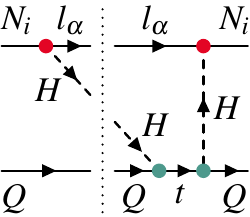} \hskip1.5mm+\hskip1.5mm\includegraphics[scale=1, valign=c]{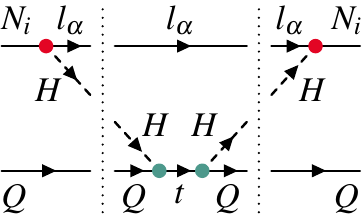}.
\end{align}
Again, using Eq. \eqref{eq9} in the first two terms, we obtain \cite{Racker:2018tzw}
\begin{align}\label{eq27}
\mathring{\gamma}_{N_iQ\rightarrow lHQ}=& 6\mathcal{Y}^2_t(\mathcal{Y}^\dagger\mathcal{Y})_{ii} \int[d\mathbf{p}_{N_i}][d\mathbf{p}_{Q}]\mathring{f}_{N_i}\mathring{f}_Q\\
&\times\frac{1}{4\pi^2}\int d^4k\delta_+[(M_i-k^0)^2-\vert\mathbf{k}\vert^2](M^2_i-k^2) \mathcal{G}_+(k,p_Q)\;2\delta_+(k^2)\mathrm{P.V.}\frac{1}{k^2}\nonumber
\end{align}
where we define
\begin{align}\label{eq28}
\mathcal{G}_\pm(k,p_Q)=\frac{2k.p_Q\pm 2m^2}{2k.p_Q\pm k^2}.
\end{align}
Note that in Eq. \eqref{eq27}, $p_Q$ is considered in the $N_i$ rest frame. Using Eq. \eqref{eq11}, the integration over $k$ gives
\begin{align}\label{eq29}
\frac{1}{4\pi}\bigg[-1-\frac{s}{M^2_i}+ \frac{M^2_i}{s-M^2_i}\left(\ln\frac{M_i\sqrt{s}}{s-M^2_i}+\ln\frac{m}{\sqrt{s}}\right)\bigg]
\end{align}
and we may immediately see that the sum of the rates in Eqs. \eqref{eq24} and \eqref{eq27} is infrared finite. The divergent logarithms in Eqs. \eqref{eq25} and \eqref{eq29} completely cancel, as discussed in Ref. \cite{Racker:2018tzw} using standard Cutkosky rules. This cancelation is typical for zero-tem\-pe\-ra\-tu\-re calculations and is based on the Kinoshita-Lee-Na\-uen\-berg theorem \cite{Kinoshita:1962ur, Lee:1964is}. It is a consequence of the stronger version of this theorem, introduced in Ref. \cite{Frye:2018xjj}, that for a forward diagram, the sum of all its cuts is infrared finite up to a forward scattering contribution. In Ref. \cite{Racker:2018tzw}, however, the diagrams with crossed quark legs, such as in Figs. \ref{fig3b} and \ref{fig3c}, were not considered. For us, these are crucial, as, without them, we would not obtain the correct form of the mass-derivative relation in this case.

\subsection{Infrared finiteness and $2\rightarrow 3$ processes at zero and finite temperature}

\begin{figure*}[t!]
\subfloat{\label{fig4a}}
\subfloat{\label{fig4b}}
\subfloat{\label{fig4c}}
\subfloat{\label{fig4d}}
\subfloat{\label{fig4e}}
\centering\includegraphics[scale=1]{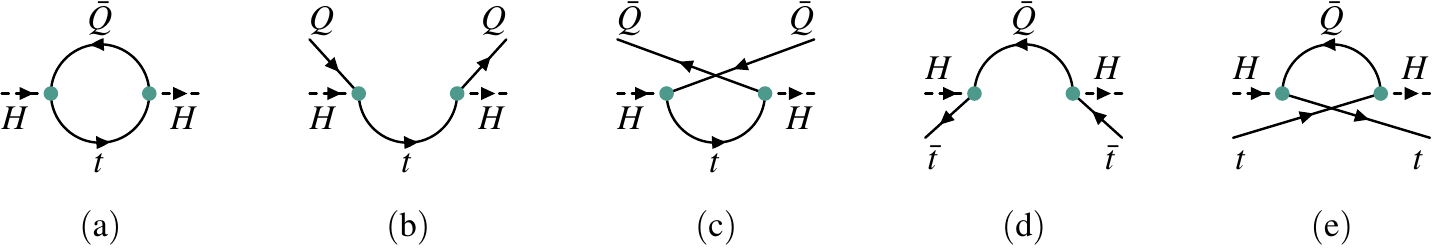}
\caption{\label{fig4} The $\mathcal{O}(\mathcal{Y}^2_t)$ zero-temperature (Fig. \ref{fig4a}) and thermal part (Figs. \ref{fig4b}-\ref{fig4e}) of the Higgs particle's self-energy to the first order in the winding number expansion \cite{Blazek:2021zoj}.}
\end{figure*}

We want to focus on the thermal mass effects and their diagrammatic representation. To understand the role of the $2\rightarrow 3$ reaction rates in this respect, let us first investigate  the $\mathcal{O}(\mathcal{Y}^2_t)$ medium contribution to the scalar self-energy. At this order, the zero-temperature self-energy of the Higgs field is equal to the diagram in Fig. \ref{fig4a} times the imaginary unit. In a thermal medium, the on-shell parts of the $Q$ and $t$ propagators are modified. We expand the phase-space densities as
\begin{align}\label{eq30}
f^{\vphantom{w}}_Q = \sum^\infty_{w=1} (\mathring{f}_Q)^w\quad\text{for}\quad\mathring{f}^{\vphantom{w}}_Q=\frac{1-f_Q}{f_Q}
\end{align}
and similarly for $t$ and the antiquark distributions. These circled densities may be understood as auxiliary functions that allow us to rewrite thermal-corrected rates as an infinite series of the zero-temperature ones \cite{Blazek:2021zoj}. To the first order in circled densities, the medium part of the self-energy denoted $\mathring{\Pi}_T$ comes from the sum of the diagrams in Figs. \ref{fig4b}-\ref{fig4d} where each external quark or antiquark is considered with the corresponding density integrated over the phase-space\footnote{Similar diagrammatic representation within the imaginary time formalism of thermal field theory can be found in Ref. \cite{Wong:2000hq}.}. Assuming all quarks and antiquarks are in thermal equilibrium, for a four-momentum $q$, each circled density has the same form $\mathring{f}_q=\exp\left\{-q^0/T\right\}$ and
\begin{align}\label{eq31}
\mathring{\Pi}_T(k^0,\mathbf{k})=6\mathcal{Y}^2_t\int[d\mathbf{q}]\:\mathring{f}_q\:
[\mathcal{G}_+(k,q)+\mathcal{G}_-(k,q)].
\end{align}
In analogy to the standard definition in Ref. \cite{Salvio:2011sf}, we introduce the thermal Higgs mass
\begin{align}\label{eq32}
\mathring{m}^2_{H,\mathcal{Y}_t}(T)=\mathring{\Pi}_T(k^0,\mathbf{k})\vert_{k^0=\vert\mathbf{k}\vert}=12\mathcal{Y}^2_t\int[d\mathbf{q}]\:\mathring{f}_q.
\end{align}
For the thermal part of the wave-function renormalization factor of the Higgs field, we have\footnote{If dimensional regularization is used instead of the finite quark mass, $\delta\mathring{Z}_{H,\mathcal{Y}_t}(T)$ vanishes in agreement with Ref. \cite{Salvio:2011sf}, as can be seen from Eq. \eqref{eq34} for $m^2\rightarrow 0$.}
\begin{align}\label{eq33}
\delta\mathring{Z}_{H,\mathcal{Y}_t}(T)=\frac{1}{2\vert\mathbf{k}\vert}\frac{\partial}{\partial k^0}\bigg\vert_{k^0=\vert\mathbf{k}\vert}\Re\mathring{\Pi}_T(k^0,\mathbf{k})
\end{align}
that multiplies the $N_i\rightarrow lH$ decay rate leading to
\begin{align}\label{eq34}
\delta\mathring\gamma_{N_i\rightarrow lH\vphantom{\bar{l}\bar{H}}}=&
6\mathcal{Y}^2_t(\mathcal{Y}^\dagger\mathcal{Y})_{ii}\int[d\mathbf{p}_{N_i}][d\mathbf{q}]\mathring{f}_{N_i}\mathring{f}_q\\
&\times -\frac{1}{4\pi^2}\int d^3\mathbf{k}\delta\left(\vert\mathbf{k}\vert-\frac{M_i}{2}\right)\frac{m^2}{(k.q)^2}\bigg\vert_{k^0=\vert\mathbf{k}\vert}
\nonumber
\end{align}
where the expression in the second row equals $1/\pi$. Furthermore, the sum of all $2\rightarrow 3$ reaction rates can be evaluated similarly to Eq. \eqref{eq27} with $\mathcal{G}_+(k,p_Q)$ replaced by $2\mathcal{G}_+(k,q)+2\mathcal{G}_-(k,q)$ coming from Figs. \ref{fig3a}, \ref{fig3d} and \ref{fig3b}, \ref{fig3c}, respectively. Unlike the integration in Eq. \eqref{eq27}, they together lead to an infrared-finite result and, remarkably, we obtain
\begin{align}\label{eq35}
\mathring{\gamma}_{N_iQ\rightarrow lHQ\vphantom{\bar{Q}}} + \mathring{\gamma}_{N_i\bar{Q}\rightarrow lH\bar{Q}} + \mathring{\gamma}_{N_it\rightarrow lHt\vphantom{\bar{Q}}} + \mathring{\gamma}_{N_i\bar{t}\rightarrow lH\bar{t}\vphantom{\bar{Q}}}
=\mathring{m}^2_{H,\mathcal{Y}_t}(T)\frac{\partial}{\partial m^2_H} \bigg\vert_{m^2_H=0}\mathring{\gamma}_{N_i\rightarrow lH} 
+\delta\mathring\gamma_{N_i\rightarrow lH}.
\end{align}
Here the two terms on the right-hand side are of equal size and represent both the effect of thermal mass and the thermal part of the wave-function renormalization of the Higgs field. The left-hand side corresponds to zero-tem\-pe\-ra\-tu\-re reaction rates obtained from specific cuttings of the forward diagrams in Fig. \ref{fig3}. As in Ref. \cite{Racker:2018tzw}, we have seen the cancelations between the $2\rightarrow 2$ and $2\rightarrow 3$ processes. Therefore, even for the reaction rates calculated using the Maxwell-Boltzmann distributions and zero-temperature Feynman ru\-les, the thermal mass effects, in terms of the expansion of Eq. \eqref{eq30}, have to be included.

Finally, one may avoid the expansion in Eq. \eqref{eq30} and work with uncircled quantum distributions corresponding to the mean occupation numbers in the Fock space. Then, the thermal mass and wave-function renormalization will appear exactly as in Refs. \cite{Giudice:2003jh,Salvio:2011sf}. Here we note that they can be represented by the finite-temperature $2\rightarrow 3$ rates obtained from cutting the forward diagrams with all possible windings of internal and external lines. That corresponds to the inclusion of the appropriate statistical factors, as it is in Eq. \eqref{eq18}. Still, the form of the mass-derivative relation in Eq. \eqref{eq35} is preserved even for uncircled quantities.

\section{Conclusions}

We have introduced a diagrammatic representation of the thermal Higgs mass in the $N_i\rightarrow lH$ reaction rate within the model of the seesaw type-I leptogenesis. It has been shown that along with $2\rightarrow 2$ scatterings, the thermal mass effects in the right-handed neutrino decay (of the same perturbative order) can and should be implemented even within the calculations based on the zero-temperature Feynman rules. The latter enters via the infrared-finite sum of $2\rightarrow 3$ reaction rates, although individual contributions diverge. Thus we identify an alternative cancelation of infrared divergences in addition to the one based on the Ki\-no\-shi\-ta-Lee-Nauen\-berg theorem used in Ref. \cite{Racker:2018tzw}.

\begin{acknowledgements}
We thank our colleague Fedor \v{S}imkovic for his long-term support. The authors were supported by the Slovak Ministry of Education Contract No. 0243/2021.
\end{acknowledgements}

\bibliographystyle{apsrev4-1.bst}
\bibliography{CLANOK.bib}

\end{document}